\documentclass{revtex4}
\usepackage{graphicx}
\usepackage{fancyhdr}
\usepackage{amsmath}
\usepackage{graphicx}
\usepackage{dcolumn}
\usepackage{bm}
\usepackage{amssymb}
\usepackage{epsfig}    
\usepackage{color}
\usepackage{slashed}
\usepackage{color}

\begin{document}

\title{
Radiative corrections to the Yukawa couplings in two Higgs doublet models
\footnotemark\footnotetext{This proceedings is based the research on Ref.~\cite{KKY yukawa}.}} 

%

\author{Mariko Kikuchi}
\affiliation{Department of Physics, University of Toyama, 3190 Gofuku, Toyama 930-8555, JAPAN}

\begin{abstract}
A pattern of deviations in coupling constants of Standard Model (SM)-like Higgs boson from their SM predictions indicates characteristics of an extended Higgs sector.
In particular, Yukawa coupling constants can deviate in different patterns in four types of Two Higgs Doublet Models (THDMs) with a softly-broken $Z_2$ symmetry.
We can discriminate types of THDMs by measuring the pattern of these deviations.
We calculate Yukawa coupling constants of the SM-like Higgs boson with radiative corrections in all types of Yukawa interactions in order to compare to future precision data at the International Linear Collider (ILC).
We perform numerical computations of scale factors, and evaluate differences between the Yukawa couplings in THDMs and those of the SM at the one-loop level.
We find that scale factors in different types of THDMs do not overlap each other even in the case with maximum radiative corrections if gauge couplings are different from the SM predictions large enough to be measured at the ILC.  
Therefore, in such a case, we can indirectly determine the type of the THDM at the ILC even without finding additional Higgs bosons directly. 
\end{abstract}

\maketitle

\thispagestyle{fancy}

\section{Introduction}
A Higgs boson was discovered at the LHC experiments and we have obtained a complete set of particles in the standard model (SM) ~\cite{LHC_Higgs}.
Observed properties of the Higgs boson are similar to those in the SM; e.g., the mass, strength of couplings and spin-parity.
The Higgs sector of the SM is composed of only an isospin doublet field.
However, there is no principle that only one doublet field must be present.
Namely, there are possibilities that the Higgs sector is extended.
Furthermore, data of the discovered Higgs boson can be explained by all extended Higgs models.
We were sure that the electroweak symmetry breaking was right, 
but we have not yet understood the essences and determined the shape of the Higgs sector.
On the other hand, it has been found that new physics models beyond the SM often contain an extended Higgs sector.  
Determining the shape of the Higgs sector by bottom up approach is one of the most important procedure to establish the new physics beyond the SM.

In this proceedings, we focus on the measurement of coupling constants of the SM-like Higgs boson ($h$) by using collider experiments.
In the SM, coupling constants of the Higgs boson are proportional to the masses of the interacting particles because all massive particles are given their masses by one Higgs field.
If there are additional Higgs fields, 
the proportional relation breaks except the case where some parameters are tuned.
Furthermore, properties of the Higgs sector appear in the pattern of deviations.
There is a possibility to discriminate extended Higgs models by comprehensively evaluating the pattern of deviations in the $h$ coupling constants in each model.

Within the relativity large uncertainties in the current LHC data ($\sqrt{s}=7, 8$ TeV, the integrated luminosity ($L$) is about 25 fb$^{-1}$), measured Higgs boson couplings look to be consistent with the SM predictions  ~\cite{Mori 2014}.
From 2015, the Higgs couplings will be measured more precisely at the second run of the LHC experiment with the collision energy to be 14 TeV with $L$ up to 300 fb$^{-1}$.
At the high luminosity LHC with $\sqrt{s}=14$ TeV and $L= 3000$ fb$^{-1}$, 
deviations in the $h$ couplings from the SM predictions can be tested with the expected accuracies about $5\%, 10\%$ and $5\%$ for $hWW(hZZ), hbb$ and $h\tau\tau$, respectively~\cite{ILC white,Peskin,TDR,CMS note}. 
Moreover, at the future linear collider such as the ILC with $\sqrt{s}=500$ GeV and $L=500$ fb$^{-1}$, those can be measured by $1.1\%, 1.6\%$ and $2.3\%$, respectively ~\cite{ILC white,Peskin, TDR, Euro}.
If we compare theoretical predictions with such precise coupling measurements at the ILC, we must evaluate the corresponding observables as precisely as possible including radiative corrections.

We here consider Two Higgs Doublet Models (THDMs) with a softly-broken $Z_2$ symmetry. 
The THDMs are well motivated in new physics models beyond the SM.
We impose the $Z_2$ symmetry to avoid flavor changing neutral currents (FCNCs). 
Consequently, there are four types of models which have the different structure of Yukawa interactions~\cite{THDMs, Akeroyd}.
We call them Type-I, Type-II, Type-X and Type-Y ~\cite{typeX, su_logan}.
In extended Higgs models with multi doublet structure, the electroweak rho parameter $\rho$ is exactly unity at the tree level. 
Because electroweak precision data indicate $\rho$ to be very close to unity; i.e., $\rho^{\textrm{exp}}=1.0004^{+0.0003}_{-0.0004}$ ~\cite{PDG}, multi Higgs doublet models such as THDMs seem to be natural.
In extended Higgs models which contain higher isospin representations, such as the Higgs Triplet Model (HTM) ~\cite{aa}, $\rho$ can deviate from unity at the tree level.
In the HTM, if a vacuum expectation value (VEV) for the triplet scalar field is tuned, the value of $\rho$ can be consistent with one of the electroweak precision data
\footnotemark\footnotetext{One-loop corrections to some Higgs couplings in Higgs triplet model with $\rho^{\textrm{tree}}\neq 1$ have been studied in Ref.~\cite{HTM_reno}.}. 
THDMs often appear in new physics scenarios.
For instance, the neutrinophilic model~\cite{neutrinophilic} contains two Higgs doublet fields with Type-I Yukawa interactions.
Yukawa interactions in the Minimal Supersymmetric Model (MSSM) correspond to those in Type-II.
There are radiative seesaw models~\cite{radiative seesaw, Kajiyama} with Type-X Yukawa interactions.

We calculate all Yukawa couplings with $h$ in all the types of THDMs including electroweak radiative corrections at the one-loop level
and evaluate deviations in these couplings from the predictions of the SM.
In the literature, the self coupling constant $hhh$~\cite{Hollik1} and Yukawa coupling constants~\cite{Hollik2} have been calculated at the one-loop level in the MSSM.
In THDMs with the softly-broken $Z_2$ symmetry, the gauge couplings $hVV(V=Z,W)$ and the $hhh$ coupling have also been calculated with one-loop corrections in Refs.~\cite{KOSC}.
However, all the Yukawa couplings to up-type quarks, down-type quarks and charged leptons have not been comprehensively analyzed including radiative corrections in the four types of THDMs.
Finally, we discuss how to discriminate the types of Yukawa interactions by combining theoretical predictions and precision measurements at the ILC.

\section{Two Higgs Doublet Models}
In the THDMs, there are two isospin doublet fields $\Phi_1$ and $\Phi_2$ whose field components are given as
 \begin{align}
 \Phi_i=\left(
      \begin{array}{c}
      \omega_i^+ \\
      \frac{1}{\sqrt{2}}(h_i+v_i+iz_i)
      \end{array}
      \right), \,\,\,
      (i=1,2).   
 \label{component1}
 \end{align} 
In Eq.(\ref{component1}), $v_i$ are VEVs for $\Phi_i$, which satisfy the relation $v^2\equiv v_1^2 + v_2^2 = (\sqrt{2}G_F)^{-1}$.
Scalar component fields with the same quantum number mix with each other.
Then physical five mass eigenstates (i.e., charged Higgs bosons $H^{\pm},$ a CP-odd Higgs boson $A$ and two Higgs bosons $h, H$) and unphysical three Nombu-Goldstone bosons $G^{\pm}, G^0$ appear.

In general, FCNCs can appear at the tree level in models with multi Higgs doublet because the Yukawa interaction matrix and the mass matrix of fermions cannot be diagonalized by one mixing. 
Taking into account constraints from flavor experiments, we should avoid FCNCs at the tree level.
We here assume that the model has a softly-broken $Z_2$ symmetry, so that one fermion can couple to only one kind of Higgs fields. 
If we assign the quantum number of the $Z_2$ symmetry to $\Phi_1$, $\Phi_2$, left-handed quark doublet, left-handed lepton doublet and right-handed up-type quark singlet fields as $+, -, +, + $and $-$, respectively,  
four types of Yukawa interaction shown in TABLE~\ref{yukawa_tab} appear. 
We call the four types as Type-I, Type-II, Type-X and Type-Y~\cite{typeX, su_logan}.
For example, the Yukawa interactions of the MSSM correspond to Type-II and there are radiative seesaw models whose the Yukawa interactions are Type-X. 

\begin{table}[t]
\begin{center}
{\renewcommand\arraystretch{1.2}
\begin{tabular}{|c|ccccccc|ccccccccc|}\hline
&
\multicolumn{7}{c|}{$Z_2$ charge}
&\multicolumn{9}{c|}{Mixing factor}\\
\cline{2-17}
&$\Phi_1$&$\Phi_2$&$Q_L$&$L_L$&
$u_R$&$d_R$&$e_R$
&$\xi_h^u$ &$\xi_h^d$&$\xi_h^e$&$\xi_H^u$&$\xi_H^d$&$\xi_H^e$ &$\xi_A^u$&$\xi_A^d$&$\xi_A^e$\\\hline
Type-I &$+$&
$-$&$+$&$+$&
$-$&$-$&$-$&
$\frac{\cos\alpha}{\sin\beta}$&$\frac{\cos\alpha}{\sin\beta}$&$\frac{\cos\alpha}{\sin\beta}$&$\frac{\sin\alpha}{\sin\beta}$&$\frac{\sin\alpha}{\sin\beta}$&$\frac{\sin\alpha}{\sin\beta}$&$\cot\beta$&$-\cot\beta$&$-\cot\beta$\\\hline
Type-II&$+$&
$-$&$+$&$+$&
$-$
&$+$&$+$
&$\frac{\cos\alpha}{\sin\beta}$&$-\frac{\sin\alpha}{\cos\beta}$&$-\frac{\sin\alpha}{\cos\beta}$&$\frac{\sin\alpha}{\sin\beta}$&$\frac{\cos\alpha}{\cos\beta}$&$\frac{\cos\alpha}{\cos\beta}$&$\cot\beta$&$\tan\beta$&$\tan\beta$\\\hline
Type-X &$+$&
$-$&$+$&$+$&
$-$
&$-$&$+$
&$\frac{\cos\alpha}{\sin\beta}$&$\frac{\cos\alpha}{\sin\beta}$&$-\frac{\sin\alpha}{\cos\beta}$&$\frac{\sin\alpha}{\sin\beta}$&$\frac{\sin\alpha}{\sin\beta}$&$\frac{\cos\alpha}{\cos\beta}$&$\cot\beta$&$-\cot\beta$&$\tan\beta$\\\hline
Type-Y &$+$&
$-$&$+$&$+$&
$-$
&$+$&$-$
&$\frac{\cos\alpha}{\sin\beta}$&$-\frac{\sin\alpha}{\cos\beta}$&$\frac{\cos\alpha}{\sin\beta}$&$\frac{\sin\alpha}{\sin\beta}$&$\frac{\cos\alpha}{\cos\beta}$&$\frac{\sin\alpha}{\sin\beta}$&$\cot\beta$&$\tan\beta$&$-\cot\beta$\\\hline
\end{tabular}}
\caption{Charge assignment of the softly-broken $Z_2$ symmetry and the mixing factors in four types of Yukawa interactions given in Eq.~(\ref{yukawa_thdm})~\cite{typeX}.}
\label{yukawa_tab}
\end{center}
\end{table}

We consider the CP conserving case in this proceedings.
Then the Higgs potential is given as
 \begin{align}
 V=&m_1^2\Phi_1^{\dagger}\Phi_1 +m_2^2\Phi_2^{\dagger}\Phi_2 
   -m_3^2(\Phi_1^{\dagger}\Phi_2+\Phi_2^{\dagger}\Phi_1)\\\nonumber
   &+\frac{\lambda_1}{2}(\Phi_1^{\dagger}\Phi_1)^2
   +\frac{\lambda_2}{2}(\Phi_2^{\dagger}\Phi_2)^2
   +\lambda_3(\Phi_1^{\dagger}\Phi_1)(\Phi_2^{\dagger}\Phi_2)
   +\lambda_4(\Phi_1^{\dagger}\Phi_2)(\Phi_2^{\dagger}\Phi_1)
   +\frac{\lambda_5}{2}\left[(\Phi_1^{\dagger}\Phi_2)^2+(\Phi_2^{\dagger}\Phi_1)^2\right],
 \end{align}
where $m_1^2,\,m_2^2,\,\lambda_1 - \lambda_4$ are real parameters, while $m_3^2$ and $\lambda_5$ are generally complex~\cite{Higgs Hunter's}.
Because we assume this model to be CP invariance, $m_3^2$ and $\lambda_5$ are real parameters.
$m_3^2$ indicates the soft breaking scale of the $Z_2$ symmetry.
Eight parameters in the Higgs potential can be rewritten into physical parameters; namely, masses of $H^{\pm}, A, H$ and $h$, two mixing angle $\alpha$ and $\beta$ which correspond to those among CP-even Higgs fields and charged (and CP-odd) Higgs fields, respectively, the VEV $v$ and the remaining parameter $m_3^2$.
We here replace $m_3^2$ by $M^2$; i.e., $M^2 = \frac{m_3^2}{\sin\beta \cos\beta}$~\cite{KOSC}.

The Lagrangian for the Yukawa interaction is shown in detail as
 \begin{align}
 \mathcal{L}^{Y}_{\textrm{THDM}} &= -\sum_{f=u, d, e} \frac{m_f}{v}
  \left(\xi_h^f \bar{f}fh + \xi_H^f \bar{f}fH -i\xi_A^f \bar{f}\gamma_5 fA\right)
    \notag\\
 & + \left[ \frac{\sqrt{2}V_{ud}}{v} \bar{u}
            \left(m_u \xi_A^u P_L + m_d \xi_A^d P_R \right)dH^+
           +\frac{\sqrt{2}m_l \xi_A^e}{v} \bar{\nu}P_R eH^+ +\textrm{h.c}
           \right],
 \label{yukawa_thdm} 
 \end{align}
where the coefficients $\xi_\varphi^f$ are summarized in TABLE~\ref{yukawa_tab}.

We here mention the coefficients of Higgs-gauge-gauge couplings $\xi_h^V (V=Z,W)$.
At the tree level, $\xi_h^V$ corresponds to $\sin(\beta-\alpha)$.
We define the SM-like limit to be $\sin(\beta-\alpha)\rightarrow 1$,
where, not only $hVV$ couplings but also $hff$ couplings approach the values in the SM. 

\section{Calculations of Radiative corrections}
In this section, we explain how to calculate one-loop corrections to the $hff$ couplings.
We here define renormalized couplings composed of three parts; i.e., tree level parts, counter-term parts and 1PI diagram parts, as follows, 
 \begin{align}
 \hat{\Gamma}_{hff} = \Gamma_{hff}^{\textrm{tree}} + \delta \Gamma_{hff}
                        + \Gamma_{hff}^{1\textrm{PI}}(p_1^2, p_2^2, q^2),
 \end{align}  
where $p_1^{\mu}$ and $p_2^{\mu}$ ($q^\mu$) indicate momentums of incoming (outgoing) particles. 

First, we shift bare parameters to renormalized parameters and counter-terms in order to obtain the formula of counter-term for the vertex.
Then the counter-term formula of Yukawa couplings are given as
 \begin{align}
 \delta \Gamma_{hff} = -i\frac{m_f}{v}\xi_{h}^f\left[
    \frac{\delta m_f}{m_f} + \delta Z_V^f +\frac{1}{2}\delta Z_h 
  + \frac{\delta \xi_h^f}{\xi_h^f} + \frac{\xi_H^f}{\xi_h^f}(\delta C_h + \delta \alpha)
  - \frac{\delta v}{v} \right],
 \end{align}
where the form of $\delta \xi_f^h$ depends on each type of Yukawa interaction.
They can be decomposed to $\delta \alpha$ and $\delta \beta$ and the exact forms are shown in Ref.~\cite{KKY yukawa}.
$\delta C_h$ is a quantity defined from Eq.(20) in Ref.~\cite{KKY yukawa}.

We here determine forms of $\delta m_f, \delta Z_V^f, \delta Z_h, \delta \alpha, \delta \beta$ and $ \delta C_h$ by imposing on-shell conditions to two point functions.
The details of these renormalization conditions are explained in Ref.~\cite{KKY yukawa}.

On the other hand, we can determine $\delta v$ from renormalization of the electroweak parameters.
In the electroweak sector of THDMs, there are five physical parameters; namely, $G_F, \alpha_{\textrm{em}}, m_Z, m_W$ and $\sin\theta_W$.
We here impose three on-shell renormalization conditions to two point functions of $ZZ$ and $WW$ and the $e\bar{e}\gamma$ vertex.
Then, we can obtain forms of $\delta m_Z^2, \delta m_W^2$ and $\delta \alpha_{\textrm{em}}$.
In this analysis, we use $\delta v$ which determined by using the relation $v^2 = 4m_W^2 \sin^2\theta_W /e^2$.
This renormalization scheme of electroweak parameters is explained in Ref.~\cite{KKY yukawa}.

Finally, we have to compute 1PI diagrams of the $hff$ vertexes.
There are sixteen kinds of diagrams in which extra Higgs bosons loop in addition to 1PI diagrams in the SM.
These complete calculations are also shown in Ref.~\cite{KKY yukawa}.

\section{Results}
In this section, we show the results of our numerical calculations.
We evaluate the renormalized scale factors defined by 
 \begin{align}
 \hat{\kappa}_f \equiv \frac{\hat{\Gamma}_{hff}(p_1^2,p_2^2,q^2)_{\textrm{THDM}}}
                            {\hat{\Gamma}_{hff}(p_1^2,p_2^2,q^2)_{\textrm{SM}}},
  \,\,\,\,\, \textrm{for}\,f=c, b, \tau 
 \end{align}
in the allowed region under constraints from perturbative unitarity and vacuum stability.
Perturbative unitarity and vacuum stability are studied in Refs.~\cite{pv_THDM} and Refs.~\cite{vs_THDM}, respectively.
We here take the external momentums to be masses of external particles; i.e., $p_1^2 = m_f^2, p_2^2 =m_f^2, q^2=m_h^2$. 
We assume that extra Higgs bosons are degenerated in the following analysis.

\begin{figure}
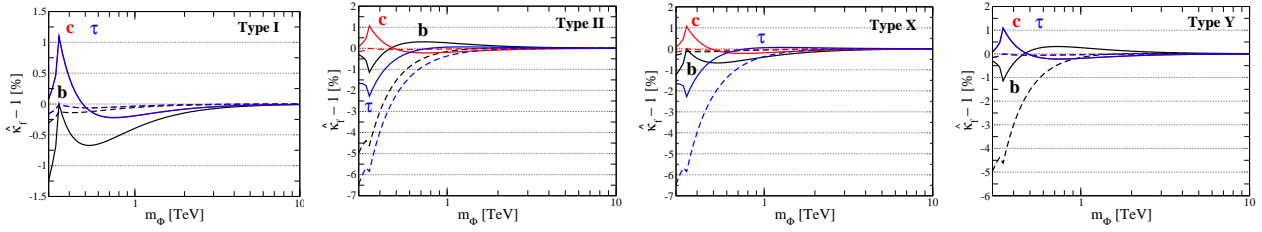

\begin{center}
\includegraphics[width=40mm]{Dec_1.eps}\hspace{1mm} 
\includegraphics[width=40mm]{Dec_2.eps}\hspace{1mm} 
\includegraphics[width=40mm]{Dec_3.eps}\hspace{1mm} 
\includegraphics[width=40mm]{Dec_4.eps}
\caption{Extra Higgs massese dependece of deviations in Yukawa couplings for $b, \tau$ and $c$ at the one loop level in the case of $\sin(\beta-\alpha)=1$. They show results in the Type-I, Type-II, Type-X, Type-Y, respectively from left to right. The value of $M^2$ is taken so as to satisfy the ralation $(300\textrm{GeV})^2=m_{\Phi}^2-M^2$. The solid (dashed) line shows results in the case with $\tan \beta = 1$ ($\tan\beta$=3).  }
\label{decouple}
\end{center}
\end{figure}

FIG. \ref{decouple} shows deviations of coupling constants of $hcc, hbb$ and $h\tau\tau$ including one-loop radiative corrections in Type-I (leftmost), Type-II (the second from the left), Type-X (the third from the left) and Type-Y (rightmost) as functions of masses of extra Higgs bosons.
We take mixing angles to be $\sin^2(\beta-\alpha)=1$ with $\tan\beta=1$ (solid line) and $\tan\beta=3$ (dashed line).
We find that the deviations can be several $\%$ even in the SM limit.
Furthermore, the patterns of the deviations among $hbb, hcc$ and $h\tau\tau$ are different in four types.
We can see that contributions of radiative corrections become to close to zero at the large mass region.
In other words, we can verify that our calculations of THDMs correspond with the predictions of the SM when extra Higgs bosons are heavier than the electroweak scale.
The peaks at around $m_\Phi=2m_t$ are caused by the resonance in the top quark loop contribution to $\Pi^{\textrm{1PI}}_{ZA}(p^2=m_A^2)$ which appears from the $\delta \beta$.

\begin{figure}
\begin{center}
\includegraphics[width=100mm]{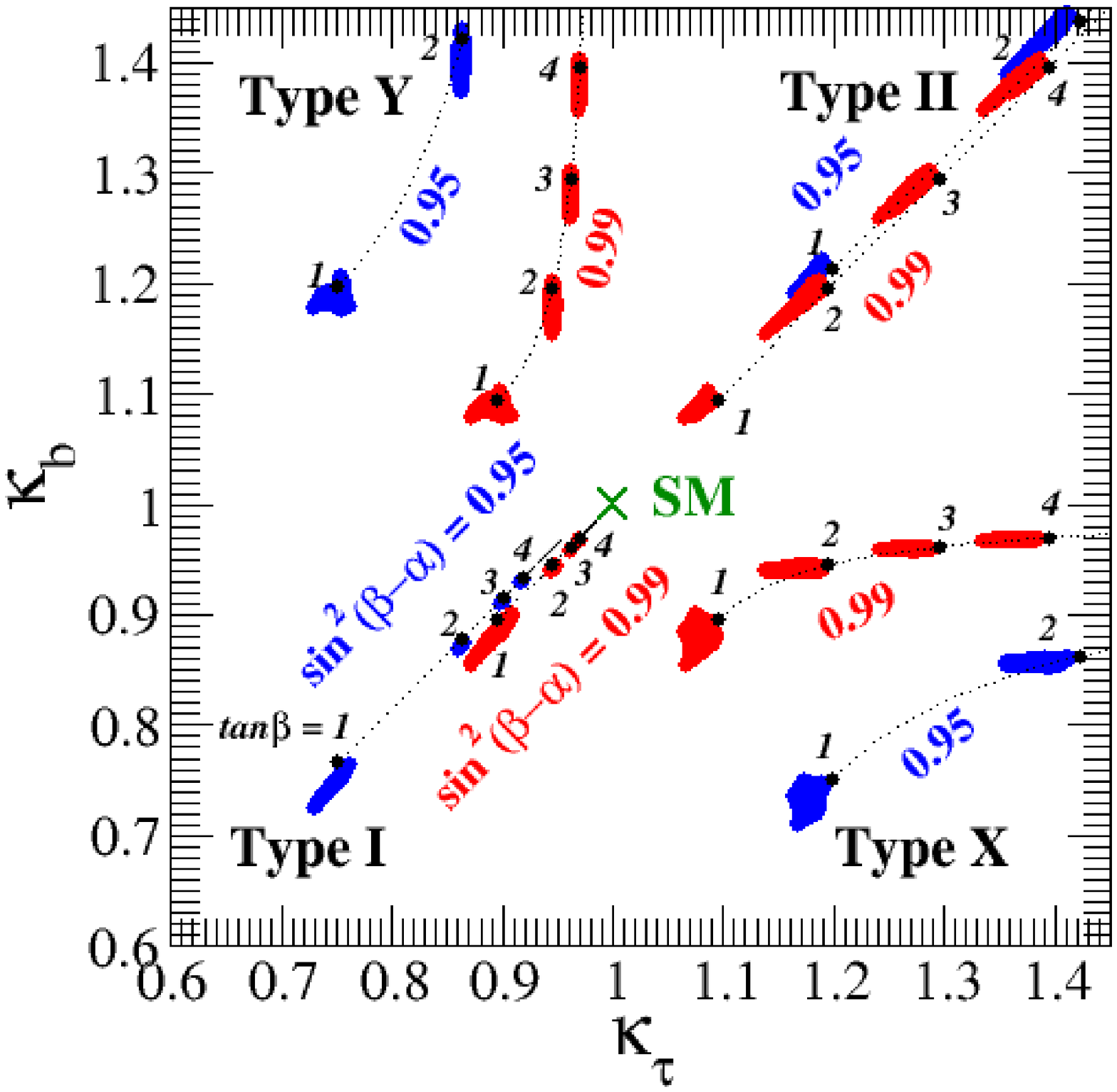}
\includegraphics[width=100mm]{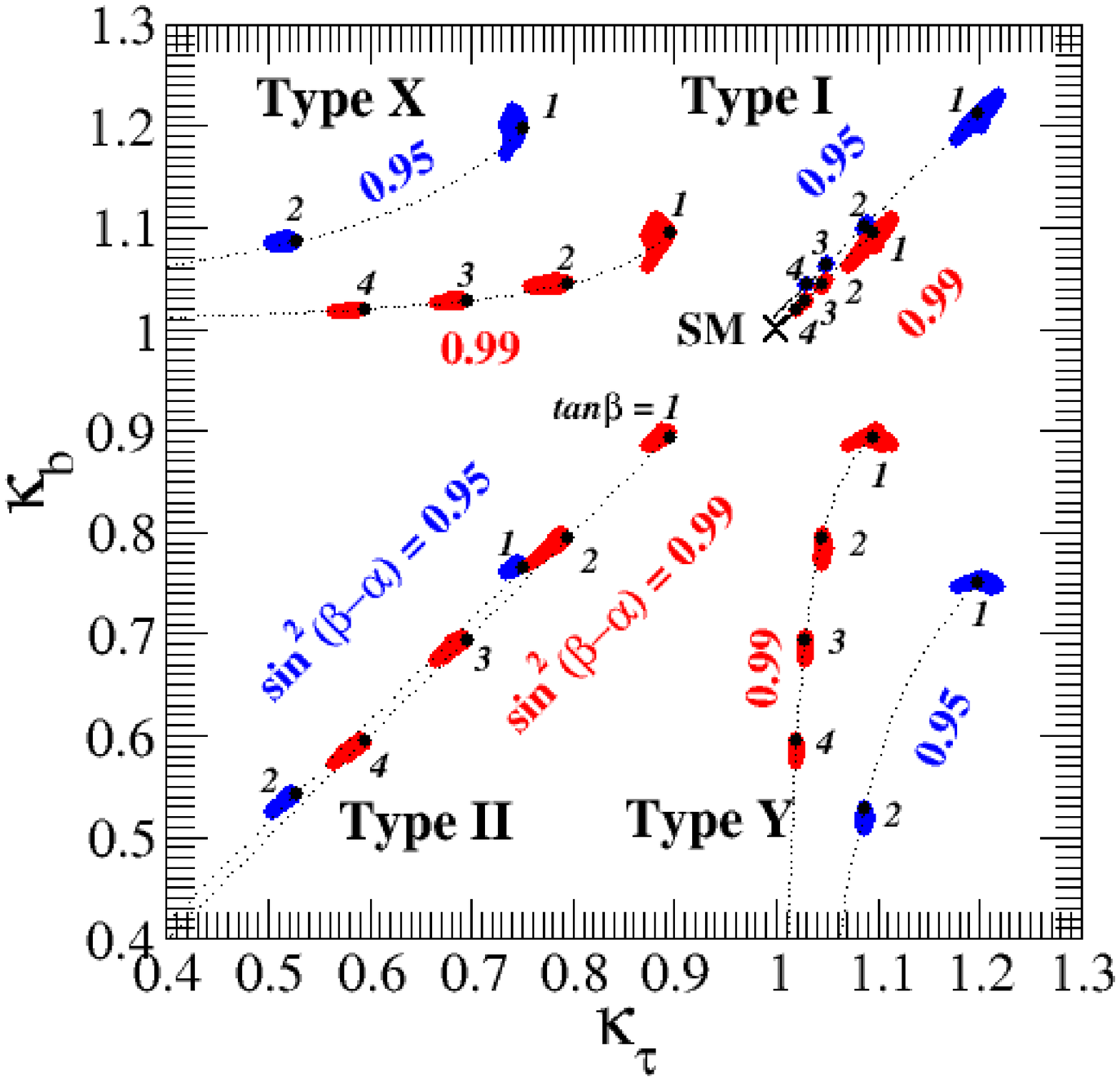}
\caption{Behavior of scale factors of $\tau$ and $b$ in four types of Yukawa interactions. The left panel and the right panel are predictions in the case with $\cos(\beta-\alpha)<0$ and $\cos(\beta-\alpha)>0$, respectively. Each black dot is a result the tree level result with $\tan\beta = 1,2,3$ and $4$. Red regions (blue regions) show one-loop results with $\sin^2(\beta-\alpha)=0.99$ ($\sin^2(\beta-\alpha)=0.95$) where $m_\Phi$ and $M$ are scanned over from 100 GeV to 1 TeV and 0 to $m_\Phi$, respectively. All predictions are favored by the constraints of perturbative unitarity and vacuum stability.} 
\label{finger1}
\end{center}
\end{figure}

In FIG. \ref{finger1}, we show the behavior of the scale factors at the tree level $\kappa_\tau^{\textrm{tree}}$ and $\kappa_b^{\textrm{tree}}$ and renormalized scale factors $\hat{\kappa}_\tau$ and $\hat{\kappa}_b$ in four types of THDMs.
The left panel and the right panel are results in the case with $\cos(\beta-\alpha)<0$ and $\cos(\beta-\alpha)>0$, respectively.
Doted lines indicate predictions at the tree level in $\sin^2(\beta-\alpha) = 0.99$ and $0.95$, and each black dots being on these lines show the tree level results with $\tan\beta = 1,2,3$ and $4$. 
At the tree level, in the case with $\sin^2(\beta-\alpha)=1$, predictions of all types approach those of the SM.
$\kappa_f^{\textrm{tree}}$ for each type lead to deviate in different directions by the situation where $\sin^2(\beta-\alpha)$ deviates from unity.
Then we can discriminate all types of Yukawa interactions by the pattern of deviations.
These analysis at the tree level have already been discussed in Refs.~\cite{ILC white, kanemura}.

We evaluate those including full electroweak and scalar bosons loop corrections which are shown by colored regions around black dots.
Red regions and blue regions are for the case with $\sin^2(\beta-\alpha)=0.99$ and $0.95$, respectively.
$m_{\Phi}$ and $M$ are scanned over from 100 GeV to 1 TeV and from 0 to $m_\Phi$, respectively.
We find that results can deviate in several $\%$ from those at the tree level due to extra Higgs loop effects. 
In the case with $m_\Phi = M$, radiative corrections become maximal by non-decoupling effects due to extra Higgs bosons loop.
Even in the case with maximal radiative corrections, predictions of $\hat{\kappa}_f (f=c,b,\tau)$ in the types of Yukawa interaction don't overlap each other, so that we can discriminate all types when $\sin^2(\beta-\alpha)$ deviates as large as $1\%$ from unity.

At the LHC with $\sqrt{s}=14$ TeV and $L=3000$ fb$^{-1}$, $h\tau\tau$ and $hbb$ couplings can be measured with about $5\%$ and $10\%$, respectively ~\cite{ILC white,Peskin,TDR,CMS note}. 
When $\sin^2(\beta-\alpha)$ is different about $1\%$ from unity, $hbb$ and $h\tau\tau$ coupling constants can deviate about $10\%$ from the predictions of the SM depending on the value of $\tan\beta$.  
In that case, we can discriminate the type of Yukawa interactions by using those high luminosity LHC data.
At the ILC, however, the Higgs coupling measurements have typically $\mathcal{O}(1)\%$ level resolution; e.g., $h$ coupling constants to $\tau$ and $b$ can be determine with $2.3\%$ and $1.6\%$ uncertainty, respectively in the version with $\sqrt{s}=500$ GeV and $L=500$ fb$^{-1}$ ~\cite{ILC white,Peskin, TDR, Euro}.
In order to compare with such precision coupling measurements at the ILC, we have to take into account the effects of radiative corrections.

\section{Conclusion}
In many new physics models, the Higgs sector is extended from the minimal one, where only one isospin SU(2) doublet is contained.
Properties of these models appear in the pattern of deviations in SM-like Higgs boson coupling constants from those of the SM.
In four types of THDMs with the softly-broken $Z_2$ symmetry, Yukawa couplings deviate from the predictions of the SM in different pattern each other, 
so that there is the possibility to discriminate all types by those correlated
 relations among Yukawa couplings.
On the other hands, it is known that $h$ coupling constants are measured typically by $\mathcal{O}(1)\%$ at the ILC.
In order to determine the Higgs sector by comparing with such high precision data, we evaluate several Yukawa couplings with radiative corrections in all types of THDMs.
We calculate loop corrections of full-electroweak sector and the scalar sector by the on-shell renormalization.
We found that each Yukawa coupling modifies about several $\%$ from the tree level prediction by extra Higgs loop corrections.
These differences are not negligible to compare with the ILC precision measurements.
If gauge couplings, such as $hWW$ and $hZZ$, slightly deviate from the SM predictions enough to measure at the ILC, 
we can distinguish all types even in the case with maximal radiative corrections.


\begin{acknowledgments}

I'm very grateful to Shinya Kanemura and Kei Yagyu for the wonderful collaborations of this work. This work was supported by JSPS, No. 25$\cdot$10031.  
\end{acknowledgments}



\bigskip 

\end{document}